\documentstyle[12pt,aasms4]{article}

\newcommand{\msun}{$M/M_{\odot}\,$}
\newcommand{\dydz}{$\Delta Y/\Delta Z\,$}
\newcommand{\bc}{$\beta$ Cephei~}

\begin{document} 

\title{Intermediate-mass star models with different helium and metal contents}

\author{Giuseppe Bono\altaffilmark{1}, Filippina Caputo\altaffilmark{1},
Santi Cassisi\altaffilmark{2,3}, Marcella Marconi\altaffilmark{4}, Luciano 
Piersanti\altaffilmark{2,5}, and Amedeo Tornamb\`e\altaffilmark{2}}

\lefthead{Bono et al.}
\righthead{Intermediate-mass stars}

\altaffiltext{1}{Osservatorio Astronomico di Roma, Via Frascati 33,
00040 Monte Porzio Catone, Italy; bono@coma.mporzio.astro.it,
caputo@coma.mporzio.astro.it}
\altaffiltext{2}{Osservatorio Astronomico di Collurania, Via M. Maggini,
64100 Teramo, Italy; cassisi@astrte.te.astro.it,
tornambe@astrte.te.astro.it}
\altaffiltext{3}{Max-Planck-Institut fur Astrophysik,
Karl-Schwarzschild-Strasse 1, 85740 Garching bei Munchen, Germany}
\altaffiltext{4}{Osservatorio Astronomico di Capodimonte, Via Moiariello 16,
80131 Napoli, Italy; marcella@na.astro.it}
\altaffiltext{5}{Universit\`a di Napoli, Dip. di Fisica, 80131 Napoli, 
Italy; piersanti@astrte.te.astro.it}

 
\begin{abstract}
We present a comprehensive theoretical investigation of the evolutionary 
properties of intermediate-mass stars.
The evolutionary sequences were computed from the Zero Age Main Sequence 
up to the central He exhaustion and often up to the phases which 
precede the carbon ignition or to the reignition of the H-shell which 
marks the beginning of the thermal pulse phase. The evolutionary tracks 
were constructed by adopting a wide range of stellar masses 
($3\leq$\msun$\leq15$) and chemical compositions. In order to account 
for current uncertainties on the He to heavy elements enrichment
ratio (\dydz), the stellar models were computed by adopting at Z=0.02 
two different He contents (Y=0.27, 0.289) and at Z=0.04 three different 
He contents (Y=0.29, 0.34, and 0.37). Moreover, to supply a homogeneous 
evolutionary scenario which accounts for young Magellanic stellar systems  
the calculations were also extended toward lower metallicities (Z=0.004, 
Z=0.01), by adopting different initial He abundances. 

We evaluated for both solar (Z=0.02) and super-metal-rich (SMR, Z=0.04) 
models the transition mass $M^{up}$ 
between the stellar structures igniting carbon and those which develop 
a full electron degeneracy inside the carbon-oxygen core. We found that 
$M^{up}$ is of the order of $7.7\pm0.5 M_\odot$ for solar composition, 
while for SMR structures an increase in the He 
content causes a decrease in $M^{up}$, and indeed it changes from 
$9.5\pm0.5 M_\odot$ at Y=0.29, to $8.7\pm0.2 M_\odot$ at Y=0.34, 
and to $7.7\pm0.2 M_{\odot}$ at Y=0.37. We also show that $M^{up}$ 
presents a nonlinear behavior with metallicity, and indeed it  
decreases when moving from Z=0.04 to $Z\approx0.001$ and increases  
at lower metal contents. This finding confirms the predictions by 
Cassisi \& Castellani (1993) and more recently by Umeda et al. (1999)
and suggests that the rate of SNe type Ia depends  on the chemical 
composition of the parent stellar population. 

This approach allows us to investigate in detail the evolutionary 
properties of classical Cepheids. In particular, we find that the 
range of stellar masses which perform the blue loop during the central 
He-burning phase narrows  when moving toward metal-rich and SMR structures. 
This evidence and the substantial decrease in the evolutionary time spent 
by these structures inside the instability strip bring out that the 
probability to detect long-period Cepheids in SMR stellar systems is 
substantially smaller than in more metal-poor systems. 

Moreover and even more importantly, we find that the time spent 
by Cepheids along the subsequent crossings of the instability strip 
also depends on the stellar mass. In fact, our models suggest 
that low-mass, metal-poor Cepheids spend a substantial portion of 
their lifetime along the blueward excursion of the blue loop, while 
at higher masses (\msun $\ge 8$) the time spent along the redward 
excursion becomes longer. Models at solar chemical composition 
present an opposite behavior i.e. the time spent along the redward 
excursion is longer than the blueward excursion among low-mass 
Cepheids and vice versa for high-mass Cepheids. 
Oddly enough, the time spent along the blueward excursion is for 
models at Z=0.01 longer than the redward excursion over the entire mass 
range. This suggests a nonlinear dependence of crossing times on 
metallicity.  
The time spent along the first crossing of the instability strip is 
generally negligible with the exception of high-mass, metal-poor stellar
structures for which it becomes of the order of 15-20\% of the total 
crossing time. 
\end{abstract}

{\em Subject headings:} stars: early-type -- evolution -- 
intermediate mass -- oscillations -- variables: Cepheids, other

\pagebreak
\section{Introduction}

A great deal of theoretical and observational studies have been  
recently devoted to the evolutionary and pulsational properties of 
metal-rich stellar structures. Thanks to the new radiative 
opacities (Seaton et al. 1994; Iglesias \& Rogers 1996, and references 
therein), it has been possible to investigate on appropriate  
physical bases the evolutionary properties of metal-rich and SMR stellar 
populations (Stothers \& Chin 1993, hereinafter SC93; Bressan, Chiosi, 
\& Fagotto 1994; Weiss, Peletier, \& Matteucci 1995; Bertelli et al. 1996; 
Salasnich et al. 1999; Girardi et al. 2000). Within this context 
Bono et al. (1997a,b,c) investigated the evolutionary properties  
of metal-rich, and SMR low-mass stars from the central H-burning 
up to the cooling sequence of white dwarfs, and provided a 
comprehensive theoretical scenario to account for the pulsation 
behavior of He-burning radial variables. 

The main aim of this paper is to extend Bono et al. (1997a,b,c) 
theoretical framework to intermediate-mass stars, i.e. to those 
stars which after burning H and He eventually form carbon-oxygen 
(CO) cores under strong electron degeneracy conditions. Due to the 
occurrence of mass loss, these stars end up their evolution as CO white 
dwarfs. Several empirical and theoretical arguments seem to support the 
evidence that CO white dwarfs belonging to close binary degenerate 
stars (Iben \& Livio 1993) become  SNe type Ia progenitors.
More massive stars succeed in igniting carbon, since they are marginally 
affected by electronic degeneracy, and therefore their final fate is to 
plausibly become the progenitor of core-collapse supernovae.  

Dating back to the pioneering investigation by 
Alcock \& Paczy\'nski (1978), the evolutionary behavior 
of intermediate  and moderately high-mass stars has been thoroughly 
investigated (Becker \& Iben 1981; Castellani, Chieffi \& 
Tornamb\'e 1983; Maeder \& Meynet 1987; Castellani, Chieffi, \& 
Straniero 1990, hereinafter CCS;  Lattanzio 1991; Bressan et al. 1993; 
Cassisi et al. 1994; Meynet et al. 1994; Castellani et al. 1999). 
However, the large majority of these predictions are based on old 
radiative opacities or typically adopt a single value of the He to 
metal enrichment ratio. In order to account for the dependence of 
intermediate-mass stars on this parameter, present calculations are  
performed by adopting for each metal abundance two or three different 
He contents, as well as the most up-to-date input physics. 

This theoretical framework will be adopted for discussing the 
evolutionary properties of two groups of variable stars, namely
\bc stars\footnote{\bc stars are a group of early-type B pulsating stars 
characterized by periods of the order of few hours, luminosity amplitudes 
which range from few hundredths to few tenths of magnitude and velocity 
amplitudes of few tens of km/sec.} and classical Cepheids. The reasons 
why we are interested in these variables is twofold:
 
1) during the last few years the observational properties of \bc stars 
have been substantially improved thanks to the high quality multiband 
Str\"omgren CCD data of young open clusters (Balona 1994; Balona \& 
Laney 1995). 
These new data play a pivotal role for providing reliable evaluations of 
both the distance modulus and the reddening of the parent cluster. 
At the same time, they also provide suitable constraints on the physical 
parameters which govern the evolutionary and pulsational properties of 
\bc stars. This occurrence makes \bc stars a useful benchmark for 
testing the observables predicted by pulsation and evolutionary 
theories for young, early-type stars. 

Detailed theoretical investigation on the evolutionary and pulsation 
properties of \bc stars have been provided by Balona, Dziembowski, 
\& Pamyatnykh (1997) and more rently by Pamyatnykh (1999).  
The evolutionary tracks presented by these authors were constructed 
by adopting fixed He abundance (Y=0.28) and two different metal 
contents, namely Z=0.02 and Z=0.03. However, empirical evidence 
based on spectroscopic measurements (Waelkens, Van den Abeele, 
\& Van Winckel 1991) suggest that the instability strip of these 
objects depends on metallicity. This finding was soundly confirmed 
by Gies \& Lambert (1992) on the basis of high signal-to-noise Reticon 
spectra of 31 field, B-type, giant stars. In fact, they found that the 
mean metallicity of their sample is of the order of Z=0.035 which means, 
without invoking error bars that the mean metallicity of field, 
early B-type stars is almost a factor of two larger than the solar 
metallicity. It is interesting to note that in the sample investigated 
by Gies \& Lambert are included 10 \bc stars and among them only 
two objects ($\xi^1$ CMa, PT Pup) are more metal-poor than the mean,  
whereas the other are more metal-rich than the mean.  
As a consequence, we are interested in implementing current evolutionary 
predictions with SMR structures. Linear, nonadiabatic pulsation 
properties of \bc stars will be addressed in a companion paper.  

2) Classical Cepheids are the most popular standard candles for 
estimating cosmic distances and recent full amplitude, nonlinear, 
convective models (Bono, Marconi, \& Stellingwerf 1999a; Bono et al. 1999b;  
Bono, Castellani, \& Marconi 2000) suggest that at fixed period 
metal-poor Cepheids are on average brighter than metal-rich ones. 
On the other hand, linear models and some observational estimates 
seem to suggest that the metallicity dependence is 
either negligible (Gieren, Fouqu\`e, \& Gomez 1998; 
Alibert et al. 1999, hereinafter ABHA) or at variance with nonlinear 
predictions (Sasselov et al. 1997; Kennicutt et al. 1998). 
This conundrum is still unsettled, but very recent empirical evidence 
suggest that Cepheid properties do depend on metallicity  
(Paczy\'nski \& Pindor 2000) and that at fixed period Galactic Cepheids 
maybe fainter than LMC Cepheids (Groenewegen \& Oudmaijer 2000).
Theoretical predictions based on both linear and nonlinear pulsation 
models rely on the Mass-Luminosity (ML) relation predicted by 
evolutionary models. To test the reliability of current ML relations 
the evolutionary framework developed in the present investigation 
will be used to evaluate the dependence of the ML relation 
on both He and metal abundances.  
Moreover in order to estimate the intrinsic spread in luminosity,  
at fixed stellar mass, we supply a detailed analysis of the time 
spent inside the instability strip during the subsequent crossings
of the pulsation region.  

The paper is organized as follows. 
In \S 2 we briefly discuss the theoretical framework adopted for 
constructing the evolutionary models, together with the selected chemical 
compositions. The main properties of the stellar evolution models are 
detailed in \S 3; H and He-burning phases are discussed 
in sections 3.1 and 3.2, while the dependence of the evolutionary behavior 
on mass loss and on $^{12}C(\alpha,\gamma)^{16}O$ nuclear reaction rate 
is presented in \S 3.3. The effects of chemical composition on $M^{up}$
are discussed in \S 3.4. Finally, the dependence of evolutionary time 
spent inside the Cepheid instability strip on chemical composition are 
presented in \S 4, together with the new ML relation. The main results 
of this investigation are summarized in \S 5. In this section we briefly
outline the observables which can help to validate the current 
evolutionary scenario.

\section{Theoretical Stellar Models}

Theoretical stellar models have been computed using the FRANEC 
(Straniero \& Chieffi 1989; Cassisi \& Salaris 1997; 
Castellani et al. 1997) evolutionary code. The OPAL radiative 
opacities (Iglesias \& Rogers 1996) are adopted for temperatures 
higher than 10,000 K, while for lower temperatures we use the 
molecular opacities by Alexander \& Ferguson (1994). Both high and 
low-temperature opacities assume a solar scaled heavy element 
distribution (Grevesse 1991). The equation of state provided 
by Straniero (1988) is supplemented at lower temperatures with 
a Saha EOS, and the outer boundary conditions are fixed according to 
the $T(\tau)$ relation by Krishna-Swamy (1966). In the outer 
layers the superadiabatic convection is treated by means of 
the canonical mixing length -{\sl ml}- formalism. We adopt 
a {\sl ml} parameter which scales with the metallicity, i.e. 
{\sl ml}=2.2 for $Z \ge 0.01$, and {\sl ml}=1.81 for Z=0.004.  
The evolutionary models at solar metallicity (Z=0.02) are  
constructed by adopting two initial He abundances, namely 
Y=0.27, and Y=0.289. The latter value is obtained by calibrating 
the present sun with a non-diffusive solar standard model 
(Salaris \& Cassisi 1996, hereinafter SC). 

It is worth mentioning that these values are in good agreement 
with recent spectroscopic measurements of M17 -a bright Galactic 
HII region- by Esteban et al. (1999) who found Y$=0.280\pm0.006$.  
To fix the chemical composition of the SMR Zero Age Main Sequence (ZAMS) 
models, a suitable assumption on the value of the \dydz parameter has to 
be provided as well. However, the value of this parameter is a thorny 
problem (Zoccali et al. 2000, and references therein), since current 
estimates are still affected by large uncertainties. 
Empirical evaluations range from \dydz$\approx6$ (Pagel et al. 1992) to
\dydz$=2.17\pm0.4$ (Peimbert, Peimbert, \& Ruiz 2000), while chemical 
evolution models seem to suggest a \dydz value equal to 1.6 for the 
solar neighborhood (Chiappini, Matteucci, \& Gratton 1997), and a 
very similar value (\dydz=1.7) for irregular galaxies 
(Carigi, Colin \& Peimbert 1999). Moreover, recent evaluations of 
the He abundance in the Galactic bulge (Renzini 1994; 
Bertelli et al. 1996) also suggest a He abundance ranging from 
Y=0.25 to 0.35. On the basis of these evidence, we adopt Y=0.34 
(\dydz=2.8) as a plausible assumption for the original He content 
of SMR stellar models (Z=0.04). However, to account for current 
uncertainties, we also compute two new sets of evolutionary tracks 
by adopting Y=0.29 (\dydz=1.5) and Y=0.37 (\dydz=3.5). 

In order to supply an evolutionary framework which covers the 
metallicity range of young stellar clusters and Cepheids in the 
Magellanic Clouds (MCs) we constructed two sets of evolutionary 
models with Z=0.01 and Z=0.004 (Luck et al. 1998). For the former 
metallicity, which is representative of the Large Magellanic Cloud 
(LMC), we adopt three different initial He contents, namely Y=0.23, 
0.255, 0.27, while for the latter, which is representative of the Small 
Magellanic Cloud (SMC), we adopt Y=0.23 and Y=0.27. 
We note that current empirical estimates suggest for the SMC a He 
content of the order of Y=0.24 (Peimbert \& Peimbert 2000). 
To properly cover the range of stellar masses which perform a "blue 
loop" in the HR diagram, the evolutionary models are constructed by 
adopting a fine mass resolution. The evolution is followed from the 
ZAMS up to the central He exhaustion and often up to the phases which 
precede carbon ignition or to the reignition of the H shell which marks 
the start of the thermal pulse phase. 

Theoretical predictions on the evolutionary behavior of intermediate 
and high-mass stars are still hampered by the treatment adopted to 
estimate, during central H-burning phase, the convective core 
overshooting beyond the formal boundary of the convective unstable 
region. Several theoretical and observational investigations have 
been focused on the evaluation of the efficiency of this mechanism 
in real stars (Chiosi et al. 1989; Chiosi, Bertelli \& Bressan 1992). 
The present theoretical scenario seems to suggest that its efficiency 
could be low (Stothers \& Chin 1992),
and therefore we fix the boundary of the convective core according 
to the canonical Schwarzschild criterion. 
The evolution of massive stars depends also on the method adopted for 
treating the semiconvective layers which develop around the H core close 
to the H exhaustion. In fact, as originally suggested by Schwarzschild \& 
H\"arm (1958) and Sakashita \& Hayashi (1959), the occurrence of such a 
phenomenon is mainly due to the electron scattering opacity and to the 
large radiative flux. Outside the Schwarzschild boundary of the 
convective core, a portion of the H-rich envelope becomes convectively 
unstable, but if the core grows, the external layers are again brought 
back to stability. 
As a consequence, a zone of partial mixing is established where 
the nuclear processed material is mixed with the H-rich layers 
until stability is achieved. 

Even though the effects of the Schwarzschild and the Ledoux criteria 
on the evolutionary behavior of massive stars is still debated
(Brocato \& Castellani 1993; Stothers \& Chin 1994; Canuto 2000, and 
references therein), we adopted the former one, since the stability 
criterion has a marginal effect on stellar models ranging from 3 to 
12 $M_\odot$.  
At the same time, the evolutionary properties of intermediate and 
high-mass stars crucially depend on the $^{12}C(\alpha,\gamma)^{16}O$ 
nuclear reaction rate, and indeed the central He burning lifetime 
(Chin \& Stothers 1991; Cassisi et al.  1998), the carbon/oxygen ratio 
-C/O- and the carbon/oxygen core mass ($M_{CO}$) at the central He 
exhaustion (Salaris et al. 1997; Umeda et al. 1999) are affected by 
this parameter. Although this nuclear reaction rate could be somehow 
constrained on the  basis of the nuclear yields in massive stars 
(Thielemann, Nomoto \& Hashimoto 1996), 
our knowledge of this rate is quite poor due to the current 
uncertainties in chemical evolution models, in massive star 
nucleosynthesis, and in the physical mechanisms which trigger 
the SNe explosion.
In the present work, we adopt the rate for the $^{12}C(\alpha,\gamma)^{16}O$ 
derived by Caughlan et al. (1985, hereinafter CFHZ85). However, in order 
to estimate the dependence of the stellar properties on this parameter 
we computed selected evolutionary tracks by adopting the rate suggested 
by Caughlan \& Fowler (1988, hereinafter CF88). 

The evolutionary properties of massive stars are also affected by 
the efficiency of mass loss. The reader interested in the evolutionary 
effects of mass loss is referred to the thorough review by 
Chiosi \& Maeder (1986) and more recently by Salasnich et al. (1999). 
The physical mechanisms which govern the mass loss in red objects 
and in particular its dependence on metal content are not well 
understood yet. A self-consistent theoretical approach for evaluating 
the mass loss rates has been recently suggested by Schaerer 
et al. (1996) who constructed a set of evolutionary models which 
include a proper treatment of the radiative transport equation in 
the outermost layers. Even though this theoretical framework seems to 
account for the mass loss rates in the blue region of the HR diagram, 
we still lack useful insights into the dependence of mass loss on metal 
content in low temperature objects. 
As a consequence, in the present investigation the mass loss 
is neglected, but to account for its effect on the evolutionary 
behavior we construct selected sequences by including  
various semi-empirical rates available in the literature.

\section{Evolutionary Properties}

The main evolutionary properties of intermediate and high-mass stellar 
models have been extensively discussed in the literature (see \S 1), 
and in this section we address in more detail the properties of metal-rich 
and SMR models (Z=0.04) and the dependence on the initial He abundance. 
Tables 1-4 list selected evolutionary parameters for both H and He 
burning phases. Columns (1) to (9) give: the stellar mass, the mass 
of the convective core at the beginning of the H-burning phase 
($M_{cc}^H$), the mass of the He core at the exhaustion 
of central H ($M_{He}^H$), the central H-burning lifetime ($\tau_H$), 
the He core mass at the beginning of the central He-burning
($M_{He}^{He}$), the mass of the convective core at the same evolutionary 
phase ($M_{cc}^{He}$), the mass of the He core ($M_{He}$) and of the 
CO core ($M_{CO}$) at the exhaustion of central He, the central He-burning 
lifetime ($\tau_{He}$). Columns (10) and (11) list the mass of the CO core 
and the surface luminosity at the second dredge up, while in the last two 
columns the luminosity and the effective temperature of the blue tip 
i.e. the hottest point reached by the model along the blue loop.  

Data listed in Tables 1-4 allow us to estimate the dependence of the 
central H-burning lifetime on the initial He abundance. 
Oddly enough, this variation is larger for less massive models 
-$16\div17$\%-, attains a minimum at $\sim8.0M_{\odot}$ and then start 
to increase at higher masses. This behavior is connected with the size 
of the convective core at the beginning of the central H-burning phase. 
Note that an increase in the He abundance causes a decrease in the size 
of the H exhausted region ($M_{He}^H$), at the ignition of the $3\alpha$ 
reaction. This effect is mainly due to the strong dependence of the mean 
molecular weight on the He abundance which causes an increase in the 
temperature of the stellar core. 

We also find that both the effective temperature and the luminosity 
of the blue tip at solar metallicity agree quite well with the results 
obtained by SC93 for Y=0.28 and Z=0.02. In fact, the difference  
between our predictions for 7 and 10 $M_\odot$ interpolated linearly 
at Y=0.28 and SC93 results (see their Table 3) is equal or smaller 
than 0.03 dex. This marginal difference could be due to the different 
set of molecular opacities adopted by SC93. Even though the formation of 
the blue loops still presents a non negligible sensitivity to input 
physics and to the physical assumptions adopted for constructing 
evolutionary models, this agreement suggests that in this mass range 
both temperature and luminosity excursions are quite robust predictions.

\subsection{The mass-luminosity-effective temperature relation for
H-burning stars}

To address several astrophysical problems concerning early type stars
we need reliable analytical relations connecting luminosity,  
effective temperature, and stellar mass. This relation is quite 
useful for \bc stars, since it allows us to constrain the luminosity 
at which these stars cross the \bc instability strip and also for 
comparing the mass estimates based on both evolutionary and pulsational 
models. 
A semi-empirical calibration of the mass-luminosity-effective temperature 
(MLT) relation for early-type stars at solar metallicity was provided 
by Balona (1984). This calibration was based on the stellar models 
available at that time (Becker 1981; Maeder 1981), while the zero point 
was fixed according to empirical mass determinations (Habets \& Heintze 
1981) and to the absolute magnitude calibration provided by 
Balona \& Shobbrook (1984).

To compare our theoretical predictions for intermediate-mass stars with 
Balona's empirical calibration, we evaluate the MLT relation for 
stellar structures  with $0.01 \le Z \le 0.04$. The relations are 
derived over the entire mass range and for two 
different evolutionary phases, namely the ZAMS and the reddest point 
along the evolutionary track before the overall contraction (OC) phase. 
The analytical relations we obtain and the uncertainties on the 
coefficients are listed in Table 5.  
Interestingly enough, we find that the stellar masses obtained by adopting 
the Balona's semi-empirical calibration are only $\approx$ 7\% higher than 
the stellar masses provided by our relation at solar metallicity. 
Although our solar metallicity models are on average $\approx 0.15$
dex fainter and $\approx 0.02$ dex cooler than the Becker's models,
the zero point adopted by Balona is supported, within the errors,
by the two sets of stellar models.

\subsection{He-burning evolutionary phases and the blue loops}

A detailed analysis of the dependence of the blue loop on the input 
physics and physical assumptions for intermediate-mass stars were 
thoroughly discussed in a seminal paper by Iben (1972) and 
more recently by Bertelli, Bressan, \& Chiosi (1985), Brunish \& 
Becker (1990), Stothers \& Chin (1991), and SC93. 
In the following we outline the main outcomes of present models.  
Figures 1, 2, and 3 show the evolutionary tracks in the HR diagram for 
the selected metallicities Z=0.004, 0.01, and Z=0.02, and the labeled
initial He contents. 
Data plotted in these figures clearly show that an increase in the 
He causes a larger excursion toward higher effective temperatures. 
This suggests that the color width between evolved blue and red 
stars could be adopted as a He indicator. 
This effect is due to the fact that an increase in the He content causes 
a decrease in the mean opacity and an increase in the mean molecular 
weight of the envelope (Vemury \& Stothers 1978), and in turn higher 
luminosities, hotter effective temperatures, and shorter H and He 
lifetimes.
However, this behavior is not linear with metallicity, and 
indeed for Z=0.01 the blue loop presents  a negligible dependence on 
He content when moving from Y=0.23 to Y=0.27 (see Figure 2). 
A quantitative explanation of this effect was not found, however 
the physical input parameters which trigger and govern the occurrence 
of the blue loops are manifold (SC93) and beyond the aim of this 
investigation.  
 
The top panel of Figure 1 shows, as expected, that an increase in 
the stellar mass widens the temperature excursion of the blue loop.  
The only exception to this behavior is the model at $5M_\odot$ 
which shows a smaller blue loop when compared with the model at 
$4.5 M_\odot$. This feature was already noted by CCS and they suggested 
that it is caused by the fact that only the stars with $M \le 4.5 M_\odot$ 
undergo a more efficient first dredge up during the RGB evolution. 
This occurrence 
causes an increase in the He content of the envelope, and in turn a 
decrease in the envelope opacity.  As a consequence, even though the He 
core mass of the $4.5 M_\odot$ model is smaller than for the $5 M_\odot$,  
the blue loop of the former model is larger.  
The same outcome applies for models at Z=0.01 -see top panel of Figure 2-, 
but the stellar mass who marks the models which experience a sizable 
first dredge up moves at \msun $\approx 5.5\div6.0$ and 
\msun $\approx 6.0$ for Y=0.255 and Y=0.27 respectively. 
Data plotted in Figures 3 and 4 do not show this feature. A detailed 
check of H-shell burning phases in metal-rich structures (Z$>0.01$) 
suggests that the amount of He dredged up during the sinking of the 
convective envelope does not show, in contrast with less metal-rich 
models, a local minimum around 5-6 $M_\odot$. This evidence suggests 
that the efficiency of the first dredge up in metal-rich structures 
-$4 \le M/M_\odot\le8$- is mainly governed by opacity and mean 
molecular weight.  

Figure 4 shows the evolution in the HR diagram of SMR models for  
three different initial He abundances (see labeled values). As expected, 
these tracks show that the excursion of the blue loop are substantially 
smaller than in more metal-poor stars. For initial He abundances of Y=0.29 
and Y=0.34, only models with stellar masses
ranging from 7.0 to 12.0 $M_\odot$ show a blue loop which crosses the 
instability strip. This mass range is further reduced to 
$7\div10 M_{\odot}$ for models at Y=0.37. 
In order to address in more detail the effect of the initial He
abundance on SMR structures, Figure 5 shows the behavior in the 
HR diagram of three models at 5, 7 and $9 M_\odot$, constructed 
by adopting the same metal content -Z=0.04- and three different initial 
He abundances. Data plotted in this figure 
show that an increase in the He abundance of 10\% ($\approx0.03$)
causes a systematic increase in $\Delta \log L$ of approximately 
0.1 dex both during H and He-burning phases. An increase in the He 
content from Y=0.29 to 0.37 causes, in contrast with more metal-poor 
stars, a decrease in the temperature excursion of the blue loop.  
The extension of the loop is also connected with the balance between the 
envelope thermal time scale (i.e. the Kelvin-Helmotz time scale) and 
the nuclear time scale of the internal structure. Therefore, 
the decrease in the temperature excursion could be due to the decrease 
in the He-burning lifetime, since the increase in the He content causes 
an increase in He core mass, and in turn a decrease in $\tau_{He}$.  
On the basis of this evidence, intermediate-mass stars with Z $> 0.02$ 
present a much lower probability to produce Cepheid variables. In fact,  
SMR structures either do not perform the blue loop (M $< 7 M_{\odot}$) 
or spend a short amount of time inside the instability strip 
(M$\ge 9 M_{\odot}$). 

Figure 6 shows the change of the central temperature as a function 
of the central density for selected stellar models. 
Note that the increase in the initial He abundance from 
Y=0.29 to Y=0.37 significantly affects the luminosity of the models,
whereas the central properties of the stellar models are only marginally 
affected at least up to the central He exhaustion. 
The origin of this behavior can be understood in terms of the quite strong 
dependence of the H-burning luminosity on the mean molecular weight which 
presents a substantial increase at higher He contents. 
Data plotted in this figure show that the thermal properties of 
intermediate-mass stars present, toward the end of the central He-burning 
phase, a strong dependence on the initial He content. This effect can 
have a crucial impact on the final fate of these structures, since  
they can be forced either to develop a CO core under conditions of 
moderate or strong electron degeneracy, or to burn carbon (see \S 3.4).

\subsection{Dependence on mass loss and on 
$^{12}C(\alpha,\gamma)^{16}O$ nuclear reaction rate}

To estimate how the mass loss affects the appearance of the blue loop,  
we computed some additional models by taking into account different 
mass loss rates available in the literature. 
Figure 7 shows the evolutionary tracks for the $8.0 M_{\odot}$, Y=0.37, 
Z=0.04 
model computed by adopting the mass loss parametrizations suggested by 
Nieuwenhuijzen \& de Jager (1990) and by Reimers (1975) respectively. 
To mimic the effects of a strong mass loss rate, the Reimers' relation 
was adopted with a large value of the free parameter $\eta$, i.e. 
$\eta=3$. Note that the typical value for low-mass, metal-poor stars 
ranges approximately from $\eta=0.3$ to $\eta=0.5$. The comparison 
between evolutionary tracks
which include or neglect the mass loss shows quite clearly that the mass 
loss supplied by the Nieuwenhuijzen \& de Jager (1990) relation slightly 
reduces 
both the excursion toward higher effective temperatures and the luminosity 
tickness of the blue loop. This finding confirms previous results for 
more metal-poor structures by Chin \& Stothers (1991) and 
Salasnich et al. (1999). On the other hand, the evolutionary track 
constructed by adopting the Reimers relation does not show at all the 
blue loop. Therefore an efficient mass loss could reduce or inhibit, 
as originally suggested by Lauterborn, Refsdal, \& Roth (1971) and by 
Lauterborn, \& Siquig (1974), the excursion toward the blue, and 
thus it cannot force metal-rich, intermediate-mass stars inside the 
instability strip. 

Figures 5 and 7 shows the occurrence of small secondary loops during 
the redward excursion of the blue loop. This feature is due to small 
increases in the size of the convective core which, in turn, causes 
small variations in the efficiency of both H-shell and central 
He-burning. This phenomenon occurs when the core He abundance is lower 
than $Y_c\approx0.3$ and it is the analog of the breathing pulses 
which take place in low-mass stars. 
The "canonical" breathing pulses become a common feature of central 
He-burning evolutionary phases when the central He abundance is lower 
than $\approx0.1$. However, their appearance is presently inhibited in 
our evolutionary code. On the contrary, the phenomenon described above 
takes place at earlier evolutionary phases and was not artificially 
quenched. This not withstanding some models constructed by inhibiting 
the appearance of such a phenomenon present only negligible differences 
in both central and He-shell burning lifetimes.

As already mentioned in \S 1, the $^{12}C(\alpha,\gamma)^{16}O$ nuclear 
reaction rate is a key physical ingredient in the evolutionary behavior 
of  intermediate-mass stars. In fact, the $M_{CO}$ 
at the He exhaustion and the C/O ratio inside this core strongly 
depends on this reaction rate. Even though a thorough analysis  
of its effects is beyond the aim of this investigation, we are 
interested in testing the dependence of our results on the 
adopted reaction rate. To account for this effect we construct 
selected models by adopting the $^{12}C(\alpha,\gamma)^{16}O$ 
reaction rate provided by CF88. 
It is worth mentioning that for temperatures typical of He-core burning 
phase, i.e. $T\approx 3\times10^8$ K, the reaction rate provided by 
CFHZ85 is approximately 2.35 larger than the CF88 one. Even though 
this reaction rate is still affected by a large uncertainty (CF88; 
Buchmann 1997), recent theoretical constraints based on evolution 
and nucleosynthesis in SNe type II progenitors support a 
$^{12}C(\alpha,\gamma)^{16}O$ reaction rate larger than suggested 
by CF88 (Woosley \& Weaver 1995; Hoffman et al. 1999). This is the 
reason why in our investigation the evolutionary models are 
constructed by adopting the reaction rate provided by CFHZ85.
To account for current uncertainties, we perform several numerical 
experiments by increasing the CF88 reaction rate by 1.7. 
This enhancement factor, which corresponds at 300 Kev to an 
$S-$factor of 170 Kev barns, accounts also for solar abundance 
distribution (Weaver \& Woosley 1993), but see also Nomoto \& 
Hashimoto (1988), and Umeda et al. (1999) for different assumptions.   

Figure 8 shows the location in the H-R diagram of selected
stellar models at solar chemical composition constructed by adopting 
different values of the $^{12}C(\alpha,\gamma)^{16}O$ rate.
Note that the use of different rates does not change,
in this mass range, the excursion of the blue loop. This finding 
supports the results of previous investigations (Chin \& 
Stothers 1991). 
At the same time, we are also interested in testing the dependence 
of the central He-burning lifetime ($\tau_{He}$), of the CO core mass
and C/O ratio at the He exhaustion, on this parameter. We find that 
for models at 5, 7, and 10 $M_\odot$ the $\tau_{He}$ values based  
on the CFHZ85 rate are $\approx$9, 7, and 8 \% longer than the 
$\tau_{He}$ values based on the CF88 rate (see Table 6), while  
the CO core mass at the He exhaustions in the former models are 
on average  3\% larger.
Therefore, the numerical experiments we performed seem to suggest that 
current uncertainties on the $^{12}C(\alpha,\gamma)^{16}O$ reaction rate 
marginally affect the Cepheid evolutionary properties. This evidence is 
further supported by the fact that both the time spent inside the 
instability strip and the tip of the blue loop are marginally affected 
by this input parameter. However, note that the total 
crossing time for the 5 $M_\odot$ model based on the CF88 rate 
is approximately 25\% shorter than the model based on the 
CFHZ85 rate. Further theoretical studies are necessary to 
investigate the impact of this behavior on short-period Cepheids.     
 
Obviously the C/O abundance ratio is, significantly affected by this 
nuclear reaction rate, and indeed models based on the CFHZ85 rate 
present a decrease of a factor of 3 when compared with the models 
based on the CF88 rate. Data listed in Table 6 also show that this 
effect does not depend on the stellar mass. 
Note that the C/O ratio affects the lifetime of white dwarfs (WD), since 
along the cooling sequence the gravothermal energy strongly depends 
on the chemical profile of the inner layers. Therefore, we can also 
expect that the $^{12}C(\alpha,\gamma)^{16}O$ reaction rate affects 
the evolutionary properties along the WD cooling sequence.

\subsection{Dependence of $M^{up}$ on metallicity}

We now discuss the evolutionary behavior of metal-rich structures after 
the central He exhaustion. At the end of the central He-burning the 
ultimate  fate of a star depends on its total mass. Stellar structures 
more massive than a critical value -the so-called $M_{up}$- quietly 
ignite carbon, evolve toward the subsequent evolutionary phases and 
eventually end up their evolution as core collapse supernovae. 
Stellar structures less massive than $M_{up}$ undergo a full electronic 
degeneracy inside the CO core during the AGB phase. These stars end up 
with a carbon deflagration -thermonuclear 
supernovae-, if and when the mass loss allows the CO core to become larger 
than the Chandrasekhar mass $M_{Ch}\approx1.4M_{\odot}$. The dependence 
of $M^{up}$ on the initial chemical composition has already been discussed 
by a number of authors (Becker 1981; Tornamb\`e \& Chieffi 1986, 
hereinafter TC; Castellani et al. 1990; Cassisi \& Castellani 1993). 
However, these investigations were focused on structures with 
$Z\le0.02$ and only recently extended to higher metallicities 
by Umeda et al. (1999).

By adopting the procedure suggested by TC, we find that the upper mass 
limit for carbon deflagration in SMR stars are $M^{up}=9.5\pm0.5 M_\odot$, 
$8.7\pm0.2 M_{\odot}$, and $7.7\pm0.2 M_{\odot}$ for Y=0.29, 0.34 and 
Y=0.37, respectively, while at solar chemical composition it is  
$M^{up}=7.7\pm0.5 M_{\odot}$. 
Figure 9 shows the comparison between different estimates 
of $M^{up}$ values as a function of metallicity. The solid line  
refers to $M^{up}$ values of models constructed by adopting 
Y=0.23 for $Z<0.01$, Y=0.255 for Z=0.01, Y=0.289 for Z=0.02, and 
Y=0.34 for Z=0.04. Our estimates have been implemented with the 
$M^{up}$ values for Z=$10^{-10}$ and $10^{-6}$ provided by Cassisi \& 
Castellani (1993). 

Data plotted in this figure show that for metallicities Z$\ge0.01$ 
our estimates agree quite well with the $M^{up}$ values estimated 
by Umeda et al. (1999). This not withstanding their $M^{up}$ value 
for Z=0.001 is approximately $1 M_\odot$ larger than our value. 
No clear explanation for this discrepancy was found, since the 
models constructed by these authors rely on update input physics. 
At the same time, they adopt a higher initial He content  
(Y=0.249 versus Y=0.23), and therefore their $M^{up}$ value 
should be smaller than our estimate. 
On the other hand, our $M^{up}$ values are systematically larger 
than the estimates provided by TC. However, predictions provided 
by TC are based on evolutionary models in which the "canonical" 
breathing pulses at the end of central He-burning were not 
inhibited. The difference between TC and our $M^{up}$ values  
confirms the results by Caputo et al. (1989), and indeed they found 
that the inclusion of breathing pulses, regardless of the stellar 
metallicity, causes an increase in the mass of CO core, and in turn 
a decrease in $M^{up}$ of $\approx0.5 M_{\odot}$. 

Finally, we note the strong dependence of $M^{up}$ on the initial He 
abundance, and indeed an increase 
in the He content from 0.34 to 0.37 causes a decrease in the $M^{up}$ 
value of the order of one solar mass. This effect is due to the increase 
in the size of the convective core during the central H-burning phase 
which, in turn causes an increase in the size of the CO core at the end 
of the central He-burning phase.

\section{The mass-luminosity relation of classical Cepheids}

The accuracy of distance determinations based on the Period-Luminosity 
(PL) and on the Period-Luminosity-Color (PLC) relations of classical 
Cepheids is widely discussed in the current literature. 
The key point in this lively debate is to assess whether Cepheids obey 
to universal PL and PLC relations or their pulsation behavior depends 
on chemical composition. Several empirical and theoretical facts such 
as the mean radius, color and the pulsation amplitudes strongly support 
the evidence that the Cepheid properties do depend on metallicity 
(Gascoigne 1974; Bono Caputo, \& Marconi 1998; 
Paczy\'nski \& Pindor 2000). Moreover, current nonlinear, 
convective pulsation models suggest, at variance with some empirical 
evidence, that metal-rich Cepheids are fainter than metal-poor ones. 

This finding relies on the adopted pulsation framework (linear vs. 
nonlinear, coupling between pulsation and convection) and on the 
ML relation adopted for constructing pulsation models. 
The dependence of the ML relation on chemical composition has been  
discussed in several papers (Chiosi, Wood, \& Capitanio 1993, 
hereinafter CWC; Saio \& Gautschy 1998; ABHA; Bono et al. 1999a; 
Bono et al. 2000). 
However, the ML relations derived by these authors rely 
on old input physics (CWC; Bono et al. 1999a) or assume at 
fixed stellar mass a mean luminosity level (Bono et al. 1999a,b; ABHA). 
To estimate the time spent inside the instability strip, we 
selected both H (1st crossing) and He-burning (2nd and 3rd crossing) 
phases located between the fundamental blue and red edges for M$>M^*$, 
and between the first overtone blue edge and the fundamental red edge 
for M$<M^*$, where $M^*$ is the predicted upper mass limit for the 
occurrence of first overtone pulsators. This limit is 7$M_{\odot}$  
for Z=0.004, 0.01 and  5$M_{\odot}$  for Z=0.02. Note that for models 
at Z=0.01 we adopted the instability edges constructed by adopting 
Y=0.25 and Z=0.008.  
The time spent inside the instability strip during the approach to 
the main sequence and after the AGB phase ($M\le M^{up}$) are 
substantially shorter 
than the crossings during H and He burning phase and have been 
neglected. 

Due to the fine mass resolution adopted for constructing the different 
sets of evolutionary models, we also investigated the minimum and the 
maximum mass which perform a blue loop 
and crosses the red edge of the instability strip. Table 7 
lists these mass values for the different chemical compositions. 
We confirm the dependence of the minimum mass on metallicity 
suggested by ABHA, and indeed we find that for \dydz=2.5 the 
$M_{min}$ changes from $\approx 3.25 M_\odot$ at Z=0.004 to 
$\approx 4.25 M_\odot$ at Z=0.01, and to $\approx 4.75 M_\odot$ 
at Z=0.02. Even though ABHA adopted slightly different He contents 
our minimum masses are quite similar to the values they predicted. 
It is worth noting that $M_{min}$ presents a similar dependence on 
He content, since an increase in He moves $M_{min}$ toward 
higher values. We also find that our upper mass limits for structures 
which perform the blue loop are for Z=0.004 and Z=0.02 roughly 
1 $M_\odot$ larger than predicted by ABHA. 
A firm explanation for this discrepancy was not found, since the 
two sets of evolutionary predictions were constructed by adopting 
the same opacity tables and the same convective instability 
criterion. However, this difference is not surprising, since as 
demonstrated by Chin \& Stothers (1991) the appearance of the blue 
loop is also affected by marginal changes in input parameters.  
  
Once the evolutionary phases which produce Cepheids are selected, 
we estimate the time spent inside the instability strip during the 
three subsequent crossings. Figure 11 shows the ratio between the 
three crossing times and the total time spent inside the instability 
strip as a function of stellar mass. 
Data plotted in this figure show quite clearly that the 
time ratios present a strong dependence on metallicity. 
In fact, the time ratios of metal-poor structures suggest 
that in the mass range $4\div6 M_\odot$ stellar structures 
spend more than 80\% of their Cepheid lifetime along the 
2nd crossing ($t^{II}$, triangles), while they spend less than 20\% 
along the 3rd one ($t^{III}$ squares) and a negligible amount of 
time along the 1st ($t^{I}$, circles). On the other hand, 
more massive structures (\msun$\ge 7$) present a different 
behavior, with $t^{II} \approx t^{III}$ for \msun=7, while  
for higher masses the time spent during the 3rd crossing 
(H and He-shell burning) becomes longer than the 2nd one 
(central He and H-shell burning). At the same time, the 
duration of the 1st crossing (H-shell burning) increases and 
becomes of the order of $15-20$\% for \msun=10-11 stellar structures. 
This suggests that for \msun=9 the probability to detected 
a Cepheid during its redward excursion (1st + 3rd crossing) is 
almost a factor of 2 larger than during its blueward excursion 
(2nd crossing). At the same time, data plotted in the top panel 
suggest that when moving from low to high-mass Cepheids the 
probability to detect a Cepheid during the 1st crossing increases
by more than one order of magnitude. 

The time ratios of stellar structures at Z=0.01 present a different 
behavior when moving from low to high-mass Cepheids. In fact, 
$t^{II}$ is generally longer than $t^{III}$ with the exception 
of the model at $M=5 M_\odot$. 
Oddly enough, the time ratios of solar metallicity models present 
an opposite behavior when compared with metal-poor structures. 
In fact in the low-mass range the time spent during the 3rd crossing 
is longer than in the 2nd one. The two ratios become identical  
for \msun=7, attain similar values up to \msun=10, but for higher 
masses $t^{II}$ is substantially longer than $t^{III}$. 
These findings suggest that in the mass range $7 \le$ \msun $\le 10$ 
the probability to detect a LMC Cepheid during its blueward 
excursion is much higher than for SMC and Galactic Cepheids. 
At the same time they also suggest that the probability to 
detect short-period Galactic and Magellanic Cepheids during 
the 1st crossing is quite negligible. 
 
The previous findings do not confirm the old rule of thumb 
that classical Cepheids are mainly evolving from the red to 
the blue (2nd crossing). In fact we find that the time spent 
during the subsequent crossings does depend not only on the 
metallicity but also on the stellar mass. We also tested 
the dependence of the three crossing times on the He content, 
and we found that it is vanishing for stellar structures at 
Z=0.01 and Z=0.02 and smaller than 10\% for Z=0.004. 
On the basis of these results and to avoid misleading 
effects  in the selection of the mean luminosity, we 
decided to derive the ML relation by including for each 
mass all the evolutionary points located inside the 
instability strip. Moreover, to properly account for the 
time spent during the three subsequent crossings, the 
individual points were weighted according to the individual 
evolutionary times. Thanks to the large range of both metallicities 
and He contents covered by our models the analytical ML relation 
was estimated by including the dependence on chemical 
composition. As a result we obtain:

\begin{tabular}{rlllll} 
$log L \,=$ & 0.90    & $+3.35\,log M$& $+1.36\,log Y$& $-0.34\,log Z$& $\;\;\;\;\;\sigma$=0.02\\
            &$\pm0.02$& $\pm0.03$     & $\pm0.13$     & $\pm0.02$     & \\
\end{tabular} 

where $\sigma$ is the standard deviation and the other symbols have 
their usual meaning. Interestingly enough we find that this ML relation 
is in very good agreement with the relation adopted by Bono et al. (1999a) 
and by ABHA. In fact, at solar composition the discrepancy with these 
relations depends on the mass value but it is always smaller than  
$\Delta \log L=0.1$. This confirms that the luminosity of intermediate-mass 
stars predicted by canonical evolutionary models is, within current 
uncertainties, well-constrained.   

Finally, we mention that the fundamental periods\footnote{In the region 
of the instability strip in which only the first overtone is unstable 
the pulsation period was fundamentalized i.e. the first overtone period 
was transformed into a fundamental period by adopting  
$\log P_F=\log P_{FO}+0.127$.} at the center of the instability strip 
provide a plain support to the empirical evidence that the period distribution 
of Magellanic Cepheids presents a sharp break in the short-period tail  
(Alcock et al. 1999; Udalski et al. 1999a,b). In fact, the OGLE samples 
clearly show that the fundamentalized period distribution of LMC Cepheids 
presents a sharp break at $\log P\approx0.4$, while in SMC Cepheids such 
a break is located at $\log P\approx0.1$. Predicted minimum fundamental 
periods for Z=0.004 and Z=0.01 attain quite similar values. In fact, 
during the 2nd and the 3rd crossing they range from $\log P\approx0.08$ 
to 0.18 for Z=0.004 and from $\log P\approx0.34$ to $\log P\approx0.44$ 
for Z=0.01.

\section{Summary and Conclusions}

We investigated the evolutionary properties of metal-rich (Z=0.01 and 
0.02) and SMR (Z=0.04) intermediate-mass stars ($3 \le$ \msun $\le 15$). 
These evolutionary calculations 
together with similar predictions for low-mass stellar structures presented 
in previous investigations (Bono et al. 1997a,b,c) supply a homogeneous 
theoretical framework for metal-rich and SMR stellar populations.   
To account for the evolutionary behavior of stellar systems in the 
MCs this theoretical scenario was implemented with two sets of 
models constructed by adopting Z=0.004 and Z=0.01 respectively. 
Owing to current empirical and theoretical uncertainties on the He 
to metal enrichment ratio and to the strong dependence of stellar 
structures on He content the evolutionary calculations were 
performed by adopting at least two different He abundances.  
The evolutionary tracks were computed from the ZAMS up to the central 
He exhaustion and often up to the phases which precede the carbon 
ignition or to the beginning of the thermal pulse phase.    

We derived the mass-luminosity-effective temperature relations 
for intermediate-mass stars along the ZAMS and during the overall 
contraction phase and we find that at solar chemical composition 
the stellar masses predicted by the new relations are in good 
agreement with values given  by the semi-empirical calibration 
suggested by Balona (1984).   
Current models support the evidence that the $M^{up}$ value, 
i.e. the cut-off mass between stars that ignite carbon under 
nondegenerate conditions and stars that form a strong 
degenerate CO core, strongly depends on the metallicity. 
This finding confirms the results by Cassisi \& Castellani (1993) 
and by Umeda et al. (1999). It is worth noting that toward 
higher metallicities -$Z > 0.02$- the $M^{up}$ value increases
and at Z=0.04 becomes approximately equal to 8 $M_\odot$, the 
exact value strongly depends on the adopted initial He content.
This result is quite interesting since it could have a substantial 
impact on chemical evolution models and on the luminosity function 
of white dwarfs in SMR stellar systems.

Thanks to the wide range of chemical compositions covered by 
current evolutionary calculations it has been possible to investigate 
in detail the evolutionary properties of classical Cepheids. 
We find that the probability to detect long-period Cepheids in 
SMR stellar systems is substantially smaller than in more metal-poor 
systems. This effect is due to the fact that the range of stellar 
masses which perform the blue loop narrows when moving toward metal-rich 
and SMR structures and also because the evolutionary time spent inside 
the instability strip is shorter than for more metal-poor ones. 

We find that the Cepheid crossing times also depends on the stellar 
mass. As a matter of fact, low-mass, metal-poor Cepheids spend a 
substantial portion of their lifetime along the 2nd crossing, while 
at higher masses (\msun $\ge 8$) the 3rd crossing time becomes longer 
than the 2nd one. On the contrary, models at solar chemical composition 
present an opposite behavior i.e. the 3rd crossing time is 
longer than the 2nd one among low-mass Cepheids, whereas it 
becomes shorter among high-mass Cepheids. At the same time, 
we also find that the crossing times present a nonlinear 
dependence on metallicity, and indeed for models at Z=0.01 
the 3rd crossing time is longer than the 2nd one over the 
entire mass range. 
It is worth mentioning that the 1st crossing time is generally 
negligible with the exception of high-mass, metal-poor stars 
for which it becomes of the order of 15-20\% of the total 
crossing time. 

Finally we mention that the outcomes concerning the difference 
in the crossing times could be directly tested on empirical data. 
In fact, these results seem to suggest that at fixed period the 
spread in luminosity among SMC Cepheids should be larger than 
among LMC Cepheids. This result might partially account for the 
empirical evidence, originally pointed out by Caldwell \& Coulson 
(1986) in optical bands and by  Laney \& Stobie (1986) in NIR bands, 
that the apparent dispersions of both PL and PLC relations are  
systematically larger for SMC than for LMC Cepheids. However, a firm 
constraint on this effect cannot be provided, since the LMC presents 
a negligible tickness along the line-of-sight, whereas we see the 
the SMC almost end-on. Therefore the luminosity scatter is expected,
due to depth effects, to be larger among SMC than LMC Cepheids. 
Note that the increase at lower metal contents in the intrinsic 
dispersion of PL and PLC relations has a marginal effect on distance 
determinations. In fact, according to Bono et al. (1999b) a systematic 
shift in the luminosity level marginally affects the slope of these 
relations, once the metallicity dependence has been properly taken 
into account.  

At the same time, these findings suggest that the period changes 
caused by evolutionary effects should depend not only on the pulsation 
period but also on the chemical composition.  
Current empirical estimates seem to suggest that period changes among SMC 
Cepheids present a different behavior when compared with LMC Cepheids 
(Deasy \& Wayman 1985). Unfortunately we lack a quantitative estimate 
of this effect, and in particular of its dependence, if any, on the 
period. In the near future, photometric data collected by large scale 
surveys can supply the accuracy needed to shed new light on these still 
unsettled questions and to constrain both evolutionary and pulsational 
predictions. 

We wish to thank V. Castellani for valuable suggestions and for a 
critical reading of an early draft of this manuscript.
One of us (S.C.) warmly acknowledge for the hospitality at the MPA 
Institute (Garching) during which part of this paper was written. 
We acknowledge an anonymous referee for his/her pertinent comments 
and useful suggestions that improved the content and the readability 
of the paper.  
The sets of evolutionary models discussed in this investigation 
can be retrived from the WWW site http://gipsy.cjb.net, while the 
isochrones are available upon request to the authors. 
This research has made use of NASA's Astrophysics Data System Abstract
Service and of SIMBAD database operated at CDS, Strasbourg, France.
This work was supported by MURST -Cofin98- under the scientific 
project: "Stellar Evolution". Partial support by ASI and CNAA is 
also acknowledged.

\pagebreak

\pagebreak
\figcaption{Evolutionary tracks in the HR diagram for stellar models 
constructed by adopting a fixed metallicity -Z=0.004- and two different 
He contents namely Y=0.23 (top panel) and Y=0.27 (bottom panel). The 
tracks cover both H and He-burning phases. The stellar masses are 
labeled.}

\figcaption{Similar to Figure 1, but for models constructed by 
adopting Z=0.01 and three different He contents: Y=0.23 (top panel),
0.255 (middle panel), and 0.27 (bottom panel).}  

\figcaption{Similar to Figure 1, but for solar metallicity models at 
Y=0.27 (top panel), and Y=0.289 (bottom panel).}  

\figcaption{Similar to Figure 1, but for SMR models constructed 
by adopting three different He contents: Y=0.29 (top panel), 
0.34 (middle panel), and 0.37 (bottom panel).}  

\figcaption{Theoretical HR diagram for selected SMR models constructed by 
adopting different He contents and stellar masses (see labeled values).}

\figcaption{Evolution of central physical conditions -temperature vs 
density- during both H and He burning phases for models constructed 
by adopting a fixed stellar mass -\msun =5.0, 7.0, 9.0- and three 
different He abundances (see labeled values).}

\figcaption{Theoretical HR diagram showing a $8.0M_{\odot}$ model 
-Z=0.04, Y=0.37- computed by adopting different empirical mass loss 
rates. The solid line refers to the canonical evolutionary track 
-no mass loss-, while the dashed and the dashed-dotted line to 
evolutionary tracks which include the mass loss relations provided 
by Nieuwenhuijzen \& de Jager (1990) and by Reimers (1978) respectively. 
See text for more details.}

\figcaption{Evolution in the HR diagram of stellar structures constructed 
at fixed chemical composition and stellar mass but different values for 
the $^{12}C(\alpha,\gamma)^{16}O$ nuclear reaction rate.}  

\figcaption{The behavior of $M^{up}$ as a function of the logarithmic 
metallicity. The $M^{up}$ values at lower metal contents were estimated 
by Tornamb\'e \& Chieffi (1986) and by Cassisi \& Castellani (1993).}

\figcaption{Time ratio between the 1st (circles), the 2nd (triangles), 
and the 3rd (squares) crossing time and the total time spent inside 
the instability strip as a function of stellar mass. The three panels
refer to models constructed by adopting different chemical compositions.}  
\end{document}